# Charge de travail du personnel infirmier dans les hôpitaux – étude bibliographique

**Gharbi Mohamed [123], Di Mascolo Maria [2], Verdier Christine [3]**
[1] Calystene, 3C rue Irène Joliot Curie, 38320, Eybens, France
[2] Univ. Grenoble Alpes, CNRS, Grenoble INP*, G-SCOP, 38000 Grenoble, France
[3] Univ. Grenoble Alpes, CNRS, Grenoble INP*, LIG, 38000 Grenoble, France

mohamed.gharbi1@univ-grenoble-alpes.fr;  maria.di-mascolo@grenoble-inp.fr; christine.verdier@univ-grenoble-alpes.fr

**Résumé.** Depuis des décennies, les services hospitaliers sont confrontés au défi de garantir des soins de qualité aux patients, malgré les pressions exercées sur le personnel en raison de la surcharge de travail. Le personnel infirmier est particulièrement touché par cette réalité, avec un ratio patients/personnel infirmier si déséquilibré que cela affecte directement la qualité des soins et le bien-être des infirmiers. Cet article examine en détail la charge de travail du personnel infirmier, en mettant en lumière les différents facteurs qui influencent cette charge, notamment le type des soins dispensés, les caractéristiques des patients et du personnel infirmier, ainsi que le contexte organisationnel. De plus, différentes méthodes de calcul et d'estimation de la charge de travail sont analysées, allant des systèmes de calcul de la charge de travail aux modèles prédictifs avancés.
Enfin, une réflexion est menée sur les pistes de recherche des futurs travaux, notamment en ce qui concerne l'identification des facteurs influençant la charge de travail, la collecte et le traitement des données ainsi que la validation des modèles.

**Abstract.** Da decenni, i servizi ospedalieri affrontano la sfida di garantire cure di qualità ai pazienti nonostante le pressioni sul personale a causa del sovraccarico di lavoro. Il personale infermieristico è particolarmente colpito da questa realtà, con un rapporto pazienti/personale infermieristico così sbilanciato da influenzare direttamente la qualità delle cure e il benessere degli infermieri. Questo articolo esamina in dettaglio il carico di lavoro del personale infermieristico, mettendo in luce i diversi fattori che influenzano questo carico, tra cui il tipo di cure fornite, le caratteristiche dei pazienti e del personale infermieristico, così come il contesto organizzativo. Inoltre, vengono analizzati diversi metodi per calcolare e stimare il carico di lavoro, dai sistemi di calcolo del carico di lavoro ai modelli predittivi avanzati. Infine, viene condotta una riflessione sulle direzioni future della ricerca, in particolare per quanto riguarda l'identificazione dei fattori che influenzano il carico di lavoro, la raccolta e l'elaborazione dei dati e la validazione dei modelli.

**Mots clés :** Planification et gestion du personnel, hôpital, charge de travail infirmier, estimation, facteurs d'influence.

## Introduction

L'augmentation du nombre de patients, le manque de personnel infirmier dans certains services, ainsi que la diversité et la complexité des soins dispensés dans les services de Médecine, Chirurgie, Obstétrique et Odontologie (MCO) ont accentué le besoin de gestion efficace des ressources hospitalières, notamment en ce qui concerne la planification du personnel infirmier.

---

*Institute of Engineering Univ. Grenoble Alpes

La profession infirmière, comme celle d'autres professionnels de santé, évolue au regard des changements organisationnels dans l'offre hospitalière, la technicité des actes mais aussi l'ampleur des tâches connexes et frontalières aux soins : charge due à la dématérialisation des dossiers patients et à la normalisation des pratiques administratives ; au développement de la relation avec le patient, la famille ou les aidants ; au glissement de certaines tâches du personnel médical vers le personnel infirmier. La profession représente un chaînon essentiel entre les patients, les aides-soignants, les médecins et les familles. Pour ces raisons, la profession infirmière est particulièrement sensible aux modifications de la charge de travail de son personnel.

Le calcul de la charge de travail du personnel infirmier requiert une évaluation complexe prenant en considération plusieurs aspects essentiels qui incluent la quantité de travail à effectuer, la complexité des tâches nécessaires pour assurer des soins de qualité aux patients, ainsi que les caractéristiques propres aux patients et au personnel infirmier. Selon [Clemens-Carpiaux, 2005] et [Deymie, 2021], en France, divers systèmes sont utilisés pour évaluer cette charge, tels que les Soins Infirmiers Individualisés à la Personne Soignée (SIIPS) et le Projet de Recherche en Nursing (PRN). Le premier système mesure la charge de travail infirmier par une évaluation globale et synthétique des soins dispensés à chaque patient, offrant ainsi une vision précise de l'intensité et de la structure du travail infirmier. Quant au second, il vise à mesurer la charge de travail temporelle des soins infirmiers par patient sur une période de vingt-quatre heures. Ce système se concentre uniquement sur la mesure du temps des toutes premières catégories d'activités : les activités de soins directs et indirects, qui constituent l'essentiel des activités mesurables par journée-patient.

Dans cet article, nous proposons un état des lieux de la littérature scientifique et professionnelle concernant la problématique de la charge de travail des personnels infirmiers. Il s'agit donc d'étudier les méthodes, modèles, outils ou techniques proposés dans la littérature pour aborder les caractéristiques et mesures de la charge de travail et les déterminants de ses modifications dans le quotidien de la pratique professionnelle.

La structure de cet article est la suivante : Nous décrivons la méthode de recherche bibliographique adoptée Ensuite, nous présentons les principaux résultats de notre recherche, mettant en lumière les différents facteurs qui influent sur la charge de travail du personnel infirmier. Enfin, nous abordons les systèmes d'estimation de cette charge. En conclusion, nous résumons nos découvertes et proposons des pistes de recherche futures basées sur cette revue de littérature.

## 1 Méthode de recherche bibliographique

Nous nous intéressons à la prédiction de la charge de travail du personnel infirmier. Pour notre recherche bibliographique, nous nous sommes basés sur la méthode de [Whittemore et Knafl, 2005], en identifiant les mots clés, les bases de données de recherche et la démarche d'affinement pour la sélection de nos articles.

Nous avons pu sélectionner quatre mots-clés *nurses, workload, prediction et hospitals*, ce qui nous a conduit à d'autres termes tels que *patient characteristics, nurses characteristics et indirect cares*. Dans la mesure où nous ciblons les personnels infirmiers dans les services de MCO en établissement hospitalier, nous avons exclu les termes suivants : *psychiatry, epidemics, pandemics, anesthesia, operating rooms, home care and home nursing.*
La recherche des articles, a été faite jusqu'à fin janvier 2024 sur les bases de données PubMed et science direct en appliquant des connecteurs de liaison AND, OR et NOT sur les mots clés. Le thésaurus MeSH a également été consulté, sans que cela n'apporte de complément intéressant par rapport à Pubmed et ScienceDirect. Cependant, il sera certainement un point d'entrée important lorsque nous aurons affiné nos pistes de recherche.
La sélection de nos articles a commencé par l'exécution de la requête (R1) : *("nurse\*" AND workload AND hospitals AND ("patient characteristics" OR "nurses experience" OR prediction) NOT pandemics NOT epidemics NOT psychiatry NOT anesthesia NOT "operating rooms" NOT "home care" NOT "home nursing")* en ne filtrant que les titres, les mots clés et les résumés. Nous avons ensuite éliminé les doublons et examiné attentivement les résumés pour ne retenir que les articles pertinents. Le nombre final des articles a été déterminé en mettant l'accent sur le contenu traitant des facteurs influençant la charge de travail et de l'estimation de cette charge.
Le tableau ci-dessous résume le nombre d'articles obtenus à chaque étape de la recherche bibliographique.

| Requête | Base de données | Nombre d'article | Nombre d'articles final |
|---------|-----------------|------------------|-------------------------|
| R1      | PubMed          | 125              | 28                      |
|         | ScienceDirect   | 25               |                         |

*Tableau 1 : Nombre d'articles retenus*

## 2 Charge de travail du personnel infirmier

La gestion de la charge de travail du personnel infirmier est l'étude qui assure l'équilibre entre la satisfaction du patient, la qualité des soins prodigués et le respect des heures de travail afin de préserver la santé du personnel. Cependant, la surcharge de travail peut causer l'épuisement, l'absentéisme et elle a également un effet sur l'intention de démission du personnel.

Afin d'assurer un équilibre au sein des équipes, il est primordial d'avoir une estimation du nombre adéquat de personnel en fonction de la charge de travail, des soins, des programmations et des plannings. De ce fait, plusieurs pistes de recherche sont intéressantes à explorer :
   i. Les facteurs et paramètres liés aux soins, aux patients et aux personnels qui influent sur la charge de travail au sein des services hospitaliers.
   ii. Les différents systèmes utilisés pour Le calcul et l'estimation de la charge de travail au sein des hôpitaux.

D'après [Kortbeek *et al.*, 2015], la charge de travail du personnel infirmier est généralement influencée par : le type de soins, le temps passé pour les soins directement liés aux patients, les caractéristiques du personnel et les compétences nécessaires pour les soins dispensés.
Nous précisons la façon dont se manifestent ces facteurs dans les paragraphes qui suivent.

**2.1 Charge de travail et type de soins**

Dans le domaine des soins, le personnel infirmier joue un rôle essentiel dans la prestation de soins de qualité aux patients. Leur charge de travail est souvent intense, variée et responsabilisante. Dans cette étude, nous explorons les différents types de soins dispensés par le personnel et leur charge de travail.
D'après [Susan *et al.*, 2006] et [Souza *et al.*, 2019], les soins prodigués par le personnel infirmier sont divisés en soins directs, soins indirects et tâches administratives, ou non-productive time, qui peuvent être incluses dans les soins indirects selon l'organisation de la structure.
Les soins directs constituent tous les soins directement liés aux patients et/ou à leurs familles. Ils englobent les soins de base, les soins techniques et les soins relationnels. Quant aux soins indirects, ils englobent les tâches non directement liées aux patients comme : la communication entre les infirmières, l'élaboration des plannings de soins et la préparation des équipements.
Enfin les tâches administratives représentent les différentes activités en relation avec la gestion des tâches du personnel infirmier au sein des services hospitaliers et leurs activités de formation. Cette diversité des soins joue un rôle très important sur la charge de travail et l'implication du personnel infirmier. [Van den Oetelaar *et al.*, 2020] ont mené une étude dans un hôpital universitaire aux Pays-Bas sur six services dans lesquels ils ont analysé les tâches administratives afin de déterminer le ratio personnel infirmier/soins. Il en ressort une grande différence entre les services en termes de taux d'implication ; une moyenne de 32% du temps est dédié aux soins directement liés aux patients ce qui signifie que le temps consacré aux soins indirects est bien supérieur au temps lié aux soins directs. Ceci s'explique par le fait que l'hôpital est très impliqué dans la formation des étudiants infirmiers, ce qui mécaniquement augmente ce temps consacré aux soins indirects

Dans une autre étude de [Souza *et al.*, 2019], dont l'objet était l'étude de l'influence des soins indirects sur la charge de travail au Brésil, les auteurs ont déterminé que les soins directs représentent approximativement 60% du temps global contre 40% pour les soins indirects. [Miller *et al.*, 2022] ont mis en lumière dans deux autres études similaires, que les tâches administratives et les soins non liés aux patients représentent un volume conséquent d'une moyenne de 30%.

Cette étude met en évidence la présence de deux types de soins, mais souligne une prédominance des soins indirects dans le temps de travail, souvent disproportionnée par rapport à leur impact direct sur les patients. De ce fait, les auteurs cherchent des alternatives telles que la délégation des soins indirects à d'autres entités hospitalières, comme proposé par [Miller *et al*, 2022]. Cette approche pourrait ainsi libérer du temps pour se consacrer davantage aux soins directs aux patients.

## 2.2. Charge de travail et caractéristiques des patients

Plusieurs études ont montré que la charge de travail dépend des caractéristiques des patients. [Van den Oetelaar *et al.*, 2020] ont mené une étude afin de définir les caractéristiques les plus influentes en se basant sur la méthode Delphi [Boulkedid *et al.*, 2011]. Cette méthode consiste à établir un questionnaire destiné aux personnels infirmiers seniors et aux cadre infirmiers afin de pouvoir déterminer les différentes caractéristiques qui influent sur la charge de travail. Quinze caractéristiques ont été mises en avant et classées en deux catégories : la dépendance mentale, comme le traumatisme, la peur, etc. et la dépendance physique, incluant l'immobilité totale ou partielle, l'obésité, l'âge, etc.

La dépendance physique a un effet significatif sur la charge de travail, notamment en influençant la nécessité de fournir des soins de bases tels que l'habillement, la toilette, etc. Selon la revue de littérature [Huang *et al.*, 2021], il ressort qu'au Brésil l'obésité avait un impact notable sur cette charge de travail. La mesure de ce facteur a été faite par l'indice de masse corporelle, révélant qu'un IMC élevé nécessite en moyenne deux heures supplémentaires par suivi médical et requiert la présence de trois personnels infirmiers ou plus.

## 2.3. Charge de travail et personnel infirmier

D'après une étude publiée dans [OMS, 2020], la profession des infirmières a connu une croissance significative ces dernières années ; entre 2013 et 2018, le nombre de personnels infirmiers est passé de 23.2 millions à 27.9 millions. Malgré cette augmentation, la surcharge de travail pour le personnel infirmier n'a pas diminué et elle impacte négativement la qualité de leurs performances. Cette situation découle en partie de la disparité de densité du personnel infirmier, allant de 83,4 pour 10 000 habitants dans les pays à revenu élevé à seulement 0,6 pour 10 000 habitants dans les pays à faible revenu.

D'après [Sari *et al.*, 2020], [Simanjorang *et al.*, 2017] et [Tong, 2018], plusieurs facteurs influencent les performances du personnel infirmier tels que : l'organisation, la surcharge de travail, le salaire, le manque de matériel, la motivation, le travail qui fait sens, le stress et d'autres plus personnels comme l'expérience, la nationalité, le sexe et le statut marital.

Parmi les facteurs cités ci-dessus, certains contribuent positivement tandis que d'autres nuisent aux performances du personnel infirmier, notamment *la surcharge de travail, le manque de motivation et le stress et l'expérience.* En outre, l'organisation joue un rôle crucial, pouvant agir à la fois comme un atout ou un obstacle. Il est essentiel d'explorer comment ces différents facteurs interagissent pour mieux comprendre leurs effets sur la qualité des services de santé fournis par le personnel infirmier.

Cette interaction entre les différents facteurs est particulièrement évidente dans le cas de la surcharge de travail. Selon [Sari *et al.*, 2019], elle est directement liée au ratio patient/personnel infirmier. Une étude menée en Californie par [Brown *et al.*, 2021] a révélé que le nombre de patients qu'un personnel infirmier peut prendre en charge varie en fonction du service, allant de 1 patient en traumatologie et blocs opératoires jusqu'à 6 patients en post-partum. Cette surcharge a été constatée également en Indonésie, avec une moyenne de 12 patients par personnel, en plus des soins indirects et des tâches administratives. Cette surcharge a donc un impact négatif sur l'implication avec le patient, entraînant des situations d'oubli et de burnout.

Selon [Simanjorgan *et al.*, 2016], il faut aussi considérer le stress qui joue un rôle très important sur la qualité des soins et des tâches. Ce dernier se manifeste au travers plusieurs paramètres comme le climat de travail ou encore les risques professionnels. Un personnel témoigne de la manière suivante : « the profession as a nurse working in a hospital environment can affect health, among others contracting disease patients, needle stick injuries, which can lead to nurse the sick ». Cela joue sur le mental du personnel avec d'autres variables comme le salaire qui est bas comparé au travail et à l'investissement demandé, la relation avec les supérieurs qui crée de la mauvaise entente et un manque de communication au sein des équipes et des relais entre équipes.

La charge de travail du personnel infirmier est le résultat d'une interaction entre plusieurs facteurs, incluant le *type et la complexité des soins requis, les caractéristiques des patients et du personnel*, ainsi que *les particularités des services hospitaliers*. Pour mieux anticiper et gérer cette charge, un croisement et une analyse de ces paramètres permettront de développer des stratégies de gestion de la charge de travail plus précises et mieux adaptées au contexte.

L'étude de [Moghadam *et al.*, 2021] menée en Iran sur le service des soins intensifs a permis de déterminer les différents facteurs ayant un impact sur la charge de travail. Pour ce faire, les auteurs, ont utilisé le Nursing Activities Score (NAS), qui est un outil permettant de calculer la charge de travail dans les services des soins intensifs. Différents indicateurs ont été utilisés concernant les caractéristiques du patient tels que l'âge, le type de traitement et la durée de séjour et les caractéristiques du personnel infirmier comme le sexe, l'âge, le niveau d'éducation et l'expérience, le type de services des soins intensifs, ainsi que les gardes (journée, soirée et nuit). Les résultats obtenus ont montré que la charge de travail est influencée d'abord par les caractéristiques des patients et du personnel infirmier et enfin par le type de garde.

## 3  L'estimation de la charge de travail

Selon [Najafpour *et al.*, 2023], la performance des systèmes de santé repose sur les quatre principes suivants : la disponibilité, l'accessibilité, l'acceptabilité et la qualité des soins. Cette dernière est définie par la satisfaction du patient et du personnel infirmier, dont le principal facteur est l'anticipation du nombre d'heures de travail adéquat du personnel par jour [Kortbeek *et al.*, 2015].

Pour atteindre cet objectif, plusieurs études ont montré que l'application des méthodes de *Machine Learning* sur différents facteurs tels que : les caractéristiques des patients et du personnel infirmier, le type et la complexité des soins, ainsi que les particularités des services hospitaliers comme le nombre de lits disponibles peut répondre à ce besoin.

D'après les différentes recherches, la gestion de la charge de travail est un sujet très complexe et aucun système ou modèle développé ne la couvre dans son entièreté. Selon [Van den Oetelaar *et al.*, 2016], [Van den Oetelaar *et al.*, 2018], et [Van den Oetelaar *et al.*, 2020], plusieurs systèmes ont été développés pour le calcul de la charge de travail, tels que RAFAELA, un système développé à l'hôpital Oulu en Finlande dans les années 90, et NZI, développé aux Pays-Bas en 1998.

[Rauhala *et al.*, 2004] ont défini le système RAFAELA comme un système de calcul de la charge de travail. Le calcul prend en compte les différentes classes de patients, le suivi journalier des tâches du personnel et les réponses du personnel infirmier à un questionnaire d'évaluation de la complexité des soins. Ce système ne prend en compte que les soins directs liés aux patients, et il n'est utilisé que pour avoir une indication sur le taux de la charge a posteriori, sans estimation ou prédiction de la charge de travail. Quant au NZI, c'est un système qui prend en compte les soins directement liés aux patients, les soins indirects et neuf caractéristiques des patients afin d'estimer et valider la charge de travail. Le système est critiqué par le personnel infirmier qui estime que les résultats fournis ne reflètent pas la réalité car certains facteurs qui leur paraissent importants ne sont pas pris en compte, comme la nécessité d'isoler certains patients et les soins psychologiques [Van den Oetelaar *et al.*, 2020].

Dans la littérature, nous trouvons plusieurs modèles pour l'estimation de la charge de travail. [Van den Oetelaar *et al.,* 2016], [Van den Oetelaar *et al.,* 2018], et [Van den Oetelaar *et al.,* 2020], ont proposé un modèle développé aux Pays-Bas prenant en compte six services de l'hôpital et qui est basé sur le système NZI, reposant sur trois grandes étapes. La première étape développée en 2016 consiste à identifier les caractéristiques des patients. La deuxième étape en 2018 consiste à mener une recherche sur les tâches du personnel et leur implication dans les différents types de soins. Enfin, lors de la troisième étape en 2020, le croisement des deux précédentes étapes a été proposé en utilisant le modèle « Linear mixed effects model » afin d'estimer la charge de travail. Notons que ce modèle ne prend pas en compte les caractéristiques du personnel, qui ont pourtant un impact sur la charge de travail.

Bien que n'étant pas le cœur de notre étude, il existe des modèles propres à certaines situations spécifiques tels que : les pandémies, la chirurgie, les blocs opératoires et la vérification des données sources (Source Data validation SDV). Les modèles utilisés relèvent de l'apprentissage supervisé et non supervisé, du Deep Learning tels que LSTM (Long-Short Term- Memory) [Sembiring *et al.,* 2024], qui sont un type de réseaux de neurones profonds appelés réseaux de neurones récurrents (RNN) [Kong *et al.,* 2017], les séries temporelles (time series), chaînes de Markov cachées (Hidden markov models HMMs) [Hanif *et al.*, 2017], qui peuvent être pertinents dans notre problématique.

[Kortbeek *et al*, 2015] ont mené une expérience afin de pouvoir anticiper le nombre de personnels adéquat dans les blocs opératoires. Pour ce faire, ils ont étudié la charge de travail requise en se basant sur le modèle « bed census prediction model » proposé par [Kortbeek *et al.,* 2012]. Ce modèle prend en compte le nombre de lits disponibles et l'estimation de la libération de ces derniers afin de pouvoir accepter ou refuser le patient. De plus, un modèle de prédiction des deux modes d'organisation du travail des personnels infirmiers (fixes et flexibles) a été implémenté. Dans cette étude, les auteurs ont abordé la notion de personnel infirmier flexible ou volatile ; c'est un pôle de personnels infirmiers pouvant passer d'un service hospitalier à un autre afin de pouvoir contrôler la charge de travail. La combinaison de ces deux modèles a permis de fournir un système d'aide à la prise de décision complet afin de pouvoir anticiper la variabilité de la charge de travail dans les blocs opératoires.

Dans l'étude menée par [Rojas *et al.*, 2018] portant sur la prédiction de la réadmission des patients dans les services de soins intensifs, les auteurs ont développé une modèle prédictif nommé « *gradient boosted machine (GBM)* ». Ce modèle est un algorithme ensembliste qui combine plusieurs modèles d'arbres de décision [Becker *et al.*, 2023]. Il se focalise sur les cas complexes à prédire, visant ainsi à améliorer la précision globale. Dans le processus, chaque arbre de décision est entraîné séquentiellement. Chaque nouvel arbre est principalement entraîné avec les données ayant été mal classées par les arbres précédents. Les résultats obtenus ont montré que le modèle est très performant comparé aux précédents outils publiés et mentionnés dans l'article.

[Mirza *et al.*, 2023], ont mis en place un système prédictif pour estimer la charge de travail liée à la vérification des données sources. L'outil est destiné aux responsables de sites et aux associés de recherche. Le modèle utilisé est une combinaison de deux algorithmes : LSTM (Last-Short-Term-Memory) et séries temporelles (Time Series) en utilisant le processus CRISP-DM [Wirth et Hipp, 2000]. Ce processus décrit le cycle de vie d'un projet de Machine Learning allant de la compréhension et l'analyse de la problématique et des données aux déploiement de la solution.

Le modèle fournit une estimation du nombre d'heures que les responsables doivent allouer à la tâche de vérification pour les mois à venir. Pour ce faire, les données sont traitées puis transformées en utilisant des techniques de séries temporelles ; enfin elles sont codées et décodées (Encoder-Decoder) en utilisant le modèle LSTM avec une couche cachée (Bottleneck) visant à réduire la dimension des entrées.

## Conclusion

L'étude menée dans cet article est une synthèse bibliographique des travaux concernant la charge de travail du personnel infirmier dans les services hospitaliers de type MCO.

Cette recherche identifie deux principales catégories : les facteurs influençant la charge de travail et l'estimation de celle-ci.

Nous avons identifié trois types de facteurs : ceux liés au type de soins, ceux liés aux caractéristiques des patients dont le niveau de dépendance, l'âge et le poids et enfin, ceux liés aux caractéristiques du personnel infirmier. Une corrélation et une complémentarité ont été observées entre ces trois volets, qui contribuent à une estimation et une évaluation plus précises de la charge de travail.

Dans cet article, nous mettons en évidence l'existence de plusieurs systèmes conçus pour le calcul de la charge de travail du personnel infirmier, tels que RAFAELA et NZI, mais qui ne contribuent pas à l'estimation et la prévision de cette dernière.

La recherche bibliographique nous a également montré que plusieurs études ont été faites dans d'autres domaines pour la prédiction de la charge de travail en utilisant des méthodes probabilistes, de régression ainsi, que des méthodes plus évoluées comme les LSTM et les Deep Generative Models.

Ces résultats ouvrent des pistes de recherche concernant de futurs travaux en se focalisant sur les facteurs à prendre en compte, la constitution des données et leurs traitements ainsi que les modèles de prédictions. Ces pistes sont listées ci-dessous :
- L'élaboration d'un questionnaire spécifique à la France pour évaluer la charge de travail du personnel infirmier, en identifiant les facteurs les plus pertinents. Cette démarche est essentielle étant donné la diversité des contextes nationaux dans lesquels les études ont été menées, chacun présentant ses propres spécificités.
- L'établissement d'une étude comparative visant à identifier les similitudes et les différences des facteurs entre les différents pays.
- L'exploration et l'étude de la corrélation entre les divers facteurs.
- L'étude et le traitement des données afin de pouvoir les exploiter.
- La vérification de la précision des charges renseignées, éventuellement en effectuant une collecte de données directement sur le terrain.

## Références